# THE DENIS 2 MICRON SURVEY AND ITS COSMOLOGICAL APPLICATIONS

Gary A. Mamon
*Institut d'Astrophysique, 98 bis blvd Arago, Paris, F-75014 France*
also *DAEC, Observatoire de Meudon, Meudon F-92195, France*
gam@iap.fr

## ABSTRACT

Near-infrared surveys have very promising cosmological applications. The DENIS 2 micron survey, currently in camera-testing, is scheduled to begin in early 1995, and to map the entire southern sky in the $I$, $J$, and $K'$ bands. The steps for the near-homogeneous extraction of $\sim 10^5$ galaxies are outlined, some pitfalls are given, and the estimated limiting capabilities of the survey are presented. A preliminary $J$-band image illustrates these capabilities and some of the hidden pitfalls. The necessary spectroscopic followup is shown to be feasible on a reasonably short timescale.

## 1 Introduction

There are two important advantages of working in the Near Infrared (hereafter NIR), in particular at the wavelength $\lambda \simeq 2\,\mu$m: 1) Interstellar dust is nearly transparent to Near Infrared radiation (at $2.2\,\mu$m, the extinction is 11% of the extinction in $V$[1,2]), allowing one to view galaxies behind the Galactic Plane and to have a view of the stellar content of galaxies, without suffering from extinction features in these (see, *e.g.*, ref. 3). 2) The total luminosity of a galaxy near $2\,\mu$m is thought to be a better tracer of the stellar mass in that galaxy than the optical light[4] (biased by recent star formation) or the far-infrared light (biased again by recent star formation that creates and heats dust grains that emit thermally in this waveband).

The blue and optical bands offer a complex view of galaxies, obscured by dust, and biased towards star forming regions, whereas the NIR bands offer a clear view of the "backbone"[3] of galaxies: their stellar distribution. Hence, it is a pity that the astronomical imagery started in the blue band and is only now developing in the NIR!

Various efforts are being currently made to dramatically increase the current database of NIR images mapping the Galaxy and the Universe. Two such efforts stand out today: the European DENIS (DEep Near-Infrared Survey of the Southern Sky) survey[5] to map the entire southern sky (a northern followup is currently being considered) in the Gunn $i$ ($0.8\,\mu$m), $J$ ($1.25\,\mu$m) and $K'$ ($2.15\,\mu$m) bands, and the American 2MASS survey[6], scheduled to map the entire sky in the $J$, $H$ ($1.65\,\mu$m), and $K'$ bands, starting around 1997. Both projects use 1m class telescopes (ESO has given its 1m to the DENIS consortium, while 2MASS will build two 1.2m telescopes, one in each hemisphere). Their cameras are based upon NICMOS $256 \times 256$ arrays, DENIS using 3″ pixels, dithered by a microscanning mirror in RA and Dec for enhanced resolution, while 2MASS will use slightly smaller (also dithered) pixels. DENIS operates in stop and stare mode, acquiring images in its 3 bands simultaneously, thanks to dichroic mirrors. Both projects should last around 3 years. The DENIS survey is currently in camera-testing, and the actual survey is expected to begin in early 1995.

## 2  Cosmological Applications

DENIS and 2MASS are designed to detect point sources up to fluxes of order 1 mJy, *i.e.*, $K \simeq 14.5$ ($10\sigma$). This flux limit is 1000 times fainter than for IRAS, and should yield $\simeq 10^8$ stars per hemisphere. DENIS was primarily designed for *stellar* applications (nearby low-mass stars — possibly including brown dwarfs —, highly obscured regions of star formation, the structure of the Milky Way and of the Magellanic Clouds). Nevertheless, its characteristics should allow the following basic *cosmological* application: obtaining a full view of the local Universe, with $\simeq 10^5$ galaxies out to $z \simeq 0.1$ (see § 3), unhampered by extinction of our own Milky Way galaxy, and unbiased by star formation or extinction in external galaxies. If stellar mass content follows the total mass content (including dark matter), then NIR surveys should be the best way to reach the distribution of matter in the Universe.

The few thousand brightest galaxies will have sufficient angular resolution to establish their morphological types and general structure. A detailed statistical study would then be performed, comparing the morphologies with that in the optical bands. For a fainter sample, one hopes that galaxies of given morphological type and "activity" (normal, starburst, AGN) will show up at precise regions of color-color diagrams, which will then be used in turn to produce candidates of given type and activity.

The structure of the Universe can be studied from any reasonably complete 2D catalog of galaxies. One expects very roughly 100 clusters, $10^4$ loose groups of 4 or more members, 100 compact groups, and over $10^4$ binaries. A very large galaxy catalog ($\sim 10^5$ objects) is necessary to understand the multiplicity function of structures with a wide dynamic range, and the correlation of properties of such structures. There are no known such complete samples of binary galaxies published yet. A compact group sample has recently been built[7] from the COSMOS galaxy catalog, showing different properties than the members of the most popular compact group catalog[8]. Large-scale structure can be assessed through a variety of clustering statistics (2-point angular correlation functions, counts in cells and void probability functions, 2-D topological genus). The power on large scales has turned out to be larger than can be accounted for with the Cold Dark Matter spectrum of primordial density fluctuations, whether measured by the angular correlation function of blue selected galaxies (APM[9], COSMOS[10]), or by the counts in cells in far-IR selected galaxies (IRAS-QDOT[11]).

It will be of considerable interest to compare the various measures of clustering as a function of waveband (hopefully, differences will be of astrophysical origin rather than from selection effects). Conversely, rather than study structure versus waveband, one can look at color versus environment. This has not been tried yet extensively for lack of large digital multicolor databases. In view of the known morphological segregation where ellipticals tend to concentrate in dense regions[12,13], the hope is that, even in 2D, *color segregation* will provide a powerful diagnostic for morphological segregation.

Galaxy counts will be established for the first time in the $1.2\,\mu$m $J$ band, and will $I$ and $K$ counts will be normalized with better precision at bright magnitudes (in comparison with deeper counts[14]).

## 3  Capabilities

How faint will DENIS extract complete and reliable lists of galaxies? To answer this question, it is useful to separate the different steps of galaxy extraction: *detection*, *photometric accuracy*, and *star-galaxy separation*. As a first step, we have relied on sets of *simulated images*, using a general purpose, home-grown, image simulation package, in which objects are placed at known positions (usually on a square grid, where on a given row, typically 20 objects have the same total magnitude with random offsets relative to pixel centers). These simulated images incorporate the effects of a Moffatt PSF, photon and read-out noise. The galaxies are modeled as $r^{1/4}$-law bulges and exponential disks. We assume that the true images can be correctly flat-fielded.

### 3.1  Galaxy Detection

Because the nominal integrations in $J$ and $K'$ last only one second, the signal-to-noise ratio is so low that galactic disks are below 1% of the sky background. It is thus necessary to *smooth* our images before *thresholding*, and to require a minimum number of *connected pixels*. We have tried various smoothing filters, and have found that simple ones (such as the *boxcar* — a 2D top-hat —) perform as well as gaussian or more sophisticated filters[15].

Requiring highly reliable detections (versus noise), we find that the detection of face-on late-type spirals is much more difficult than for edge-on spirals or ellipticals, and requires a considerably larger smoothing radius. Of course, detection is hampered at low galactic latitude where extinction can become important, especially in the shorter wavelength $I$ band.

### 3.2  Galaxy Photometry

An extensive study of isophotal photometry shows that the magnitude offset (relative to the true total magnitude) is a function of object type and inclination, sky background, and to a lesser extent to instrumental gain. We have also tested an iterative circular aperture photometric technique[16], which performs about the same.

### 3.3 Star/Galaxy Separation

Low surface brightness galaxies, such as face-on late-type spirals, may be hard to detect, but once detected, they are easily distinguishable from stars. On the other hand, ellipticals are easy to detect, but difficult to tell apart from stars. The large pixel size and the high ratio of stars to galaxies (from 100 to over $10^4$) make star-galaxy separation a difficult task. Using a simple star/galaxy separation statistic (the mean surface brightness above a threshold within an aperture, see ref. 16), we find that star/galaxy separation is a function of object type, and of course of the level of confusion by stars[17].

Star/galaxy separation will be possible, *through the Galactic Plane* (for $|\ell| > 45°$), with a loss of 1–1.5 magnitudes, before extinction. Thanks to its better spatial

Table 1: Estimated DENIS galaxy extraction limits

| Detection at pole | $I$ | $K_I^0$ | $J$ | $K_I^0$ | $K$ | $K_I^0$ |
|---|---|---|---|---|---|---|
| Ellipticals | 17.5 | 15.8 | 16.4 | 15.4 | 14.6 | 14.6 |
| Edge-on Disks | 17.4 | 16.0 | 16.4 | 15.5 | 13.9 | 13.9 |
| Face-on Disks | 16.8 | 15.4 | 15.7 | 14.8 | 12.9 | 12.9 |
| Star/galaxy separation at pole | $I$ | $K_I^0$ | $J$ | $K_I^0$ | $K$ | $K_I^0$ |
| Ellipticals | 16.5 | 14.7 | 14.4 | 13.4 | 13.2 | 13.2 |
| Edge-on Disks | 16.7 | 15.3 | 15.6 | 14.7 | 13.4 | 13.4 |
| Face-on Disks | 17.2 | 15.8 | 15.7 | 14.8 | 13.6 | 13.6 |
| Star/galaxy separation at $\ell = 90°, b = 0°$ | $I$ | $K_I^0$ | $J$ | $K_I^0$ | $K$ | $K_I^0$ |
| Ellipticals | 14.5 | 9.3 | 14.3 | 12.3 | 12.6 | 12.2 |
| Photometry at pole | $I$ | $K_I^0$ | $J$ | $K_I^0$ | $K$ | $K_I^0$ |
| Ellipticals | | | | | 13.6 | 13.6 |
| Edge-on Disks | | | | | 13.4 | 13.4 |
| Face-on Disks | | | | | 12.4 | 12.4 |

NOTES: The detection limits are for 95% completeness, the star/galaxy separation limits are for 92% reliability, and the photometric limits are for 0.2 mag accuracy. $K^0$ refers to equivalent $K$ magnitudes, corrected for extinction.

resolution, the $I$ band should provide the best star/galaxy separation at high galactic latitudes. However, at low galactic latitudes, the $J$ and especially $K'$ bands provide better star/galaxy separation, in terms of unextinguished magnitude limits[17].

Table 1 summarizes the extraction estimates. A clean and complete $K$-selected catalog would go to $K = 12.4$, while giving up on the face-on late type spirals will allow one to reach $K = 13.4$ (where $L_*$ galaxies go out to $z = 0.09$). At these magnitude limits, there are 2.5 and 9 galaxies per square degree (see ref. 14), respectively, yielding 50 000 and 100 000 galaxies, respectively, for these two catalogs. An $I$ selected catalog would go to $I = 16.5$, yielding $\simeq 10^6$ galaxies.

## 4 A First View of DENIS Galaxies

Figure 1 shows a $J$-band DENIS image of a field (at $b = 15°$) chosen to contain the only galaxy in a given ($12'$ RA by $30°$ Dec) strip, bright enough to appear in the LEDA extragalactic database. This image was observed in the early phases of DENIS camera testing (January 1994). Recent images show considerable quality improvement (fewer ghosts, much less vignetting and negligible center-to-edge PSF degradation), but are not yet fully processed. The LEDA galaxy ($B = 14.7$, labeled

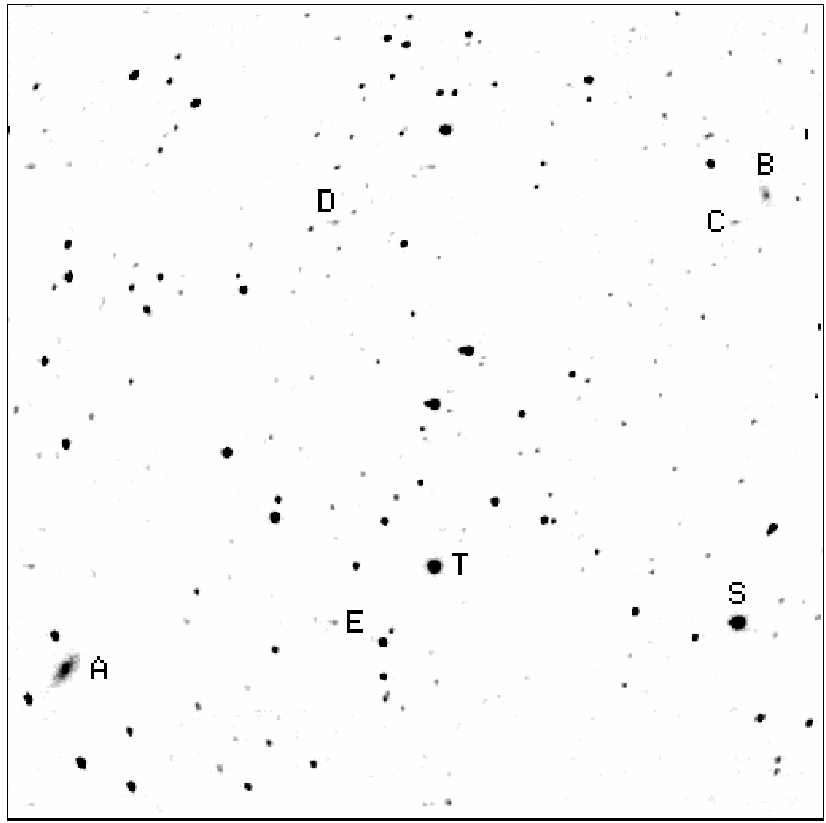

Figure 1: DENIS ($12' \times 12'$) image (average of 9 subimages) in $J$. A is the LEDA galaxy ($B = 14.7$), B is a candidate galaxy, while C is a ghost of star S, as well as D and E, and star T has ghosts too. Grey-levels span from 2.5 to $20\,\sigma$ above the background.

A on the figure) is clearly visible. Two other galaxy candidates B and C are both confirmed with our star/galaxy separation algorithm, although object C turns out to be a ghost of star S.

## 5 Steps to Ensure Optimal and Homogeneous Extraction

How can we best ensure that the galaxy catalogs extracted from the DENIS image database will be sufficiently homogeneous for cosmological applications, and yet as large as possible? One one hand, we need to optimize our *algorithm parameters*, and on the other we need to know our *selection functions* (completeness in detection, reliability in detection [versus noise], reliability in star-galaxy separation, photometric offset and dispersion), and ensure that these are as homogeneous as possible.

Our strategy is as follows: 1) We construct and test our algorithms on simulated images, fine-tuning our parameters for optimal and homogeneous extraction. 2) We take repeated exposures of certain DENIS strips at high-galactic latitude, co-add them, and build a *truth-table* of galaxies by applying our algorithms with the 1st order parameters based upon the analysis of the simulated images. Since our simulated images are idealized, our algorithms will not perform optimally at the magnitude limit. But for the co-added images, they should perform satisfactorily if we limit ourselves to the expected magnitude limit of single exposures. 3) We compare our galaxy list to a blue selected galaxy list (*e.g.,* from the APM or COSMOS surveys): indeed the blue surveys are complete to $b_J \simeq 20.5$, faint enough to include all the galaxies extracted by DENIS, unless there exists a very red class of galaxies (recall that $L_*$ galaxies are visible out to $z = 0.1$, so that $k$-corrections will be relatively small). If any bright galaxy in the blue list is missing from the NIR list, then we may have a serious problem with our algorithm. Conversely, if all galaxies from the NIR list are included in the blue list, this could mean that blue selected samples can serve as truth tables (except that we will not have good NIR photometry on these galaxies to understand our selection functions in terms of NIR magnitudes), although this could also mean that our algorithms are too inefficient. 4) We fine-tune our algorithm parameters by optimizing the extraction of galaxies from our individual exposures using our NIR truth-table. 5) We study the effects of confusion by adding to our repeated exposures at high galactic latitude stars extracted from low galactic latitude fields, and try to recover the galaxies (this was suggested to us by D. Lynden-Bell). We mimic the effects of extinction by either decreasing the pixel values of our objects or by adding an additional background (trying to be consistent with the characteristics of its noise).

## 6 Hurdles for Homogeneous Galaxy Extraction

Besides its intrinsic hurdles (large pixel size, confusion by stars) and the fact that the selection functions strongly depend on object type, there are a few additional hurdles of instrumental origin: In the image shown in Figure 1, the size of the PSF increases by a factor 3 between the optical center (in the middle of the upper right quadrant) and the lower left corner. Such variations in the PSF will have a strong influence on star/galaxy separation. Fortunately, this PSF problem was corrected very recently:

the optical center is now closer to the center, and its degradation is now no more than 20% at the corners, and usually much less.

Bright stars leave ghosts, which look very similar to galaxies in our star/galaxy separation plots. However, these ghosts can be identified by their positions, as they occur at 3 corners of a square, whose sides are parallel to the pixel grid, with the bright star lying in the fourth corner (see Fig. 1). The ghosts are much less frequent in the recent images. Dust specs also look like galaxies. They can be identified because they tend to lie at the same place during consecutive frames, although they tend to move a little. The instrumental gain is such that the background standard deviation is of the order of 1–4 ADU. This poor sampling affects our photometric accuracy, although only moderately.

## 7 Spectroscopic Followup

As for most 2D surveys, it will be important to work on a spectroscopic followup. The redshifts obtained with low spectral resolution spectra will allow one to map the local Universe in 3D out to $z \simeq 0.1$, to study structures (binaries, compact groups, loose groups, clusters, and large-scale structure), their statistics (spatial correlation functions, etc.) in 3D, and internal kinematics. Spectra with higher spectral resolution will provide distance indicators, independent of redshift, and hence will yield rough estimates of the peculiar velocities of galaxies, from which one can attempt to recover the underlying mass-density field, but this is a longer-term proposal.

Obtaining $> 10^5$ redshifts is a costly proposal, even at low spectral resolution. Table 2 shows estimates of the number of clear, dark nights required to cover completely a hemisphere with low-resolution spectroscopy. The 2dF on the AAT is the most efficient system, even at bright limiting magnitudes. Sutherland is competitive at $B_{\text{lim}} = 17$, and FLAIR (see Parker, in these proceedings) would be very competitive to $B = 18$ with a fast robotic positioner and 270 fibers. Spread over 5 years, a 2dF key-program would require 35, 50, and 85 *clear, dark* nights, annually, to $B = 17, 18$ and 19, respectively (shared with other scientific programs?), while a semi-dedicated improved FLAIR could complete most of a shallow ($B = 17$) survey in only 2 years.

I thank Paul Felenbok, Véronique Cayatte, and Catherine Boisson for discussions on the characteristics of the instruments listed in Table 2.

Table 2: Estimated timescales for a spectroscopic followup

| Telescope (instrument) | $D$ (m) | $\phi$ (deg) | $N_{\rm mos}$ | $B_{\rm crit}$ | $t_{\rm ovhd}$ (mins) | $t^{17}_{2\pi\,\rm sr}$ (nts) | $t^{18}_{2\pi\,\rm sr}$ (nts) | $t^{19}_{2\pi\,\rm sr}$ (nts) |
|---|---|---|---|---|---|---|---|---|
| UKST (FLAIR-II)   | 1.2 | 7.3 | 92  | 16.1 | 150 | 650  | 2400 | 11000 |
| UKST (FLAIR-III?) | 1.2 | 7.3 | 270 | 17.0 | 150 | 280  | 900  | 4000  |
| UKST (FLAIR-IV?)  | 1.2 | 7.3 | 270 | 17.0 | 10  | 70   | 360  | 2500  |
| Sutherland (FIFI?)| 1.9 | 2.1 | 30  | 17.2 | 10  | 290  | 1550 | 9300  |
| Las Campanas      | 2.5 | 1.0 | 150 | 19.9 | 10  | 1000 | 1600 | 3300  |
| ESO (MEFOS)       | 3.6 | 1.0 | 30  | 18.5 | 10  | 770  | 1100 | 4100  |
| AAT (2dF)         | 3.9 | 2.0 | 400 | 19.6 | 10  | 180  | 250  | 420   |
| CTIO (ARGUS)      | 4.0 | 1.0 | 24  | 18.3 | 10  | 730  | 1000 | 4200  |
| VLT (FUEGOS)      | 8.0 | 0.5 | 80  | 20.6 | 10  | 2400 | 2600 | 3300  |

NOTES: The numbers given in the table are approximate estimates, assuming that all cameras have the same optical throughput. $\phi$ is the angular diameter of the equivalent circular field of view, $t_{\rm ovhd}$, $t_{17}$, $t_{18}$, and $t_{19}$, indicate the overhead time between successive integrations, and the time to complete a hemispheric survey (without regard to the very high extinction at very low galactic latitudes) to limiting magnitudes of $B = 17$, 18, and 19, respectively. $B_{\rm crit}$ is the magnitude where all the fibers become used. FLAIR III and FLAIR IV are hypothetical instruments with increased number of fibers and, for the latter, an efficient robotic positioner. The FIFI instrument is not available on Sutherland, but could be easily mounted.